# PHOTOMETRIC STUDY OF THE UNUSUAL BINARY SYSTEM VSX J052807.9+725606

N. A. Virnina[1], I. L. Andronov[1,2], K. A. Antoniuk[2]

(1) Department "High and Applied Mathematics", Odessa National Maritime University, Odessa, Ukraine; e-mail: virnina@gmail.com, tt_ari@ukr.net
(2) Crimean Astrophysical Observatory, Crimea, Ukraine; e-mail: antoniuk@crao.crimea.ua

*The results of three-color BVR photometric study of a recently discovered unusual binary system VSX J052807.9+725606 = USNO-B1.0 1629-0064825 are presented. This system is very similar to V361 Lyr, which was assumed to be unique. We confirmed a strong asymmetry of the phase curve and found the wavelength dependence of the amplitude. This is interpreted by a "direct impact" model with a "hot spot" in the atmosphere of the accreting component. Color temperatures are determined. Characteristics of the "hot spot" are estimated. We also calculated the new ephemeris for VSX J052807.9+725606.*

**Introduction**

During more than 20 years the well studied short periodic ($P=0.309^d$) binary system V361 Lyr with an extremely asymmetric light curve figured in many papers as an unique system with a direct impact of accretion stream with atmosphere of star, which caused rather large steady hot region. According to the classical model of this star, due to the spot, the phase curve exhibits a strong asymmetry, which appears to be wavelength dependent: the amplitude of asymmetry increases with decreasing wavelength. The photometric study [1, 2, 3, 4, 5, 6, 7], spectral observations [8] and modeling [2, 3, 5] yields the interacting system consisting of two near-solar mass stars. The analysis of the *O-C* diagram [7, 8, 9] shows fast period decrease. Under the assumption of conservative mass transfer, this shows that the donor star is more massive and loses matter very rapidly. The mass stream doesn't form an accretion disk, but impacts into the photosphere of less massive and less bright star, causing a very hot region - the "direct impact" model. This assumption is in a very good agreement with the photometric and spectral studies. In 1990 J. Kaluzny [3] suggested that V361 Lyr could be a binary system, "caught" at the final stage of transition from detached to contact configuration, and called this stage "a precontact binary".

In 2010 Virnina and Andronov [10] discovered variability of the object USNO-B1.0 1629-0064825 (R.A.(2000) = $05^h28^m07.975^s$, Dec.(2000) = +72°56'06.05") and classified it as a new short period ($P=0.41179^d$) binary system, which got in the electronic catalogue "Variable Star Index" the designation VSX J052807.9+725606. The characteristics of the phase curve of this object are very similar to that ones observed in V361 Lyr.

They published unfiltered phase curve and pointed out its uncharacteristically large asymmetry of maxima and the shift of the secondary minimum from the phase 0.5. Such significant difference between magnitudes in maxima couldn't be explained by the O'Connell effect [11], which magnitude is usually smaller by one to two orders of magnitude [11, 12, 13, 8]. For the explanation of the effect, several basic models were discussed, including the periastron passage [11], the spots on one or both components of the system [14] and circumstellar gas [15].

Thus, the authors of [10] preliminary supposed that the very bright "hot spot" model, proposed for V361 Lyr in [2] and qualitatively confirmed by spectral observations [8], could be suitable also for the new variable.

**Observations**

According to Virnina and Andronov [10], VSX J0528 is a rather faint star. They observed the new variable using the remotely controlled astrophysical refractor AP180, combined with unfiltered CCD camera SBIG STL-11K. The authors noticed that the maximum quantum efficiency of the camera sensor is close to the standard R-band. Thus, using the R-band magnitudes for the comparison stars, they obtained $min_I=16.^m590\pm0.^m020$, $min_{II}=16.^m387\pm0.^m018$, and $max_I=15.^m924\pm0.^m016$ and $max_{II}=16.^m213\pm0.^m020$.

In this paper we present the results of B, V and Rc observations, performed in the Crimean Astrophysical Observatory using the 1.25 m AZT-11 telescope equipped with the CCD camera FLI PL1001E-1 during 4 nights (27 hours) from JD 2455538 to 2455625. The image resolution was 0.92"/pixel, the field of view was 7.85'x7.85'. The exposure times were $exp_B=120$, $exp_V=60$ and $exp_R=30$ seconds. Altogether we obtained 372, 395 and 386 points in B, V and R filters, respectively. The numbers of points, obtained during each observational night, are listed in the Table 1.

TABLE 1. Journal of observations.

| Date | B | V | R |
|---|---|---|---|
| 2010-12-07 | 109 | 116 | 114 |
| 2010-12-08 | 133 | 141 | 137 |
| 2011-01-05 | 124 | 132 | 129 |
| 2011-03-04 | 6 | 6 | 6 |
| Total | 372 | 395 | 386 |

**Calibration**

There were no well-known comparison stars in the vicinity of VSX J0528. The comparison stars, used in [10] are out of bounce of the smaller field of view of AZT-11. To obtain magnitudes of the comparison stars for the investigated variable, we used 3 stars in the vicinity of the Mira-type variable SU Cam (RA(2000) = $06^h 38^m 12.66^s$, Dec(2000) = +73°54'58.3"). We took *BVR* images of VSX J052807.9+725606 and SU Cam alternately, during one night (2011-03-04), when the fields were at nearly the same altitude, and, consequently, close air-masses. We used the information about comparison stars, given as Henden's standards on the AAVSO chart 4677edf [16]. The coordinates, AAVSO numbers, USNO-B1.0 numbers and *B, V, Rc* magnitudes of 3 comparison stars in the field of SU Cam are summarized in the Table 2. The comparison stars and SU Cam itself are marked on the Fig. 1.

TABLE 2. Information about comparison stars near SU Cam.

| AAVSO | USNO-B1.0 | RA | Dec | B | V | R |
|---|---|---|---|---|---|---|
| 142 | 1639-0055745 | $06^h 39^m 19.^s182$ | +73° 54″ 42.70′ | 14.890 | 14.159 | 13.728 |
| 146 | 1638-0055544 | $06^h 38^m 48.^s264$ | +73° 53″ 32.57′ | 15.260 | 14.617 | 14.249 |
| 151 | 1639-0055718 | $06^h 38^m 56.^s988$ | +73° 56″ 04.53′ | 15.831 | 15.081 | 14.620 |

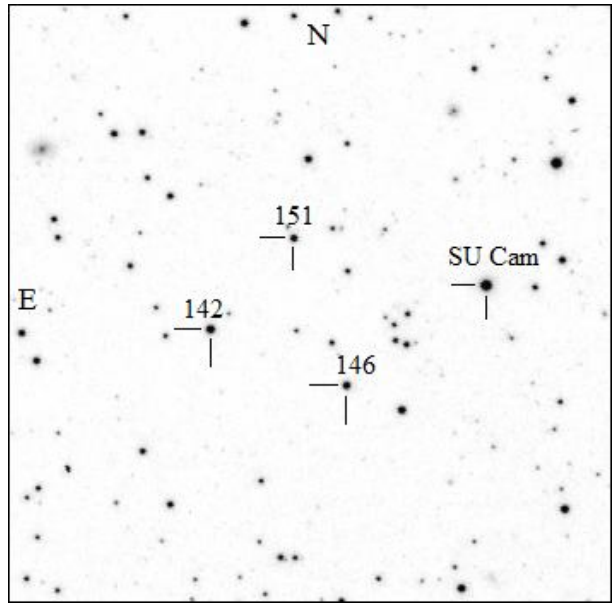

Fig. 1. Comparison stars near SU Cam.
Coordinates of center R.A. (2000)=06$^h$38$^m$55$^s$, Dec.(2000)=+73°54'58".

As the comparison stars for VSX J0528, we chose 5 stars in its vicinity, and calculated their instrumental magnitudes, using the relation of their intensities to the intensities of reference stars near SU Cam, and the classical Pogson's formula:

$$m = m_1 - 2.5 \cdot \lg\left(\frac{I}{I_1}\right),$$

where $m$ is an unknown magnitude of a given comparison star, $m_1$ - known magnitude of the comparison star near SU Cam, $I$ and $I_1$ – are their intensities in a corresponding filter.

. The chart for the chosen comparison stars is shown on the Fig. 2, while the information about the coordinates, names and mean instrumental magnitudes with corresponding error estimates is presented in the Table 3. To improve accuracy, we have used 5 comparison stars and equal weights for averaging. The corresponding r.m.s. error estimate for such an "artificial comparison star" are 0.013$^m$, 0.018$^m$ and 0.007$^m$ for instrumental photometric system $b, v, r$.

To use statistically independent signals, we have used measurements in the instrumental systems. However, for color index and temperature calibration, a conversion to the standard system is needed. According to H. Hardy [17] the magnitude difference between the reference star and the variable one, measured in the

instrumental system, could be transformed into the standard photometric system using the following equations

$$\Delta V = (1-k_v k_{bv})\Delta v + k_v k_{bv}\Delta b,$$

$$\Delta B = (k_{bv} + k_v k_{bv})\Delta b + (1 - k_{bv} - k_v k_{bv})\Delta v,$$

$$\Delta R_c = (k_{vr} + k_v k_{vr})\Delta r + (1 - k_{vr} - k_v k_{vr})\Delta v,$$

$$\Delta(B-V) = k_{bv}\Delta(b-v),$$

$$\Delta(V-R_c) = k_{vr}\Delta(v-r),$$

where $\Delta V$, $\Delta B$, $\Delta R_c$, $\Delta(B-V)$, $\Delta(V-R_c)$ are magnitude difference and colors in the standard system, $\Delta v$, $\Delta b$, $\Delta r_c$, $\Delta(b-v)$, $\Delta(v-r_c)$ are corresponding measured differences in the instrumental system, $k_v$, $k_{bv}$, $k_{vr}$ – the coefficients of reduction, determined from the observations of several tens standard stars from the Landolt's list [18].

According to K.N.Grankin (private communication), the reduction coefficients are $k_v = -0.0177$, $k_{bv} = 1.4286$, $k_{vr} = 1.0046$. Using these values of coefficients, derived by K.N.Grankin, and the formulae mentioned above, we get relations between the instrumental and standard photometric system:

$$\Delta B = 1.4033\Delta b - 0.4033\Delta v, \quad \Delta V = 1.0253\Delta v - 0.0253\Delta b, \quad \Delta Rc = 0.9868\Delta r + 0.0132\Delta v,$$

$$\Delta(B-V) = 1.4286\Delta(b-v), \quad \Delta(V-Rc) = 1.0046\Delta(v-r),$$

where $\Delta m$ – is magnitude difference between the variable star and the comparison star in a given filter. The effective wavelengths of the instrumental systems, derived from these coefficients, are $\lambda_b = 466nm$, $\lambda_v = 548nm$, $\lambda_v = 640nm$.

Using these color transformation equations, we have computed magnitudes of comparison stars in the "standard" system using an "unweighted mean" artificial comparison star near SU Cam. The corresponding values and their error estimates are listed in Table 4.

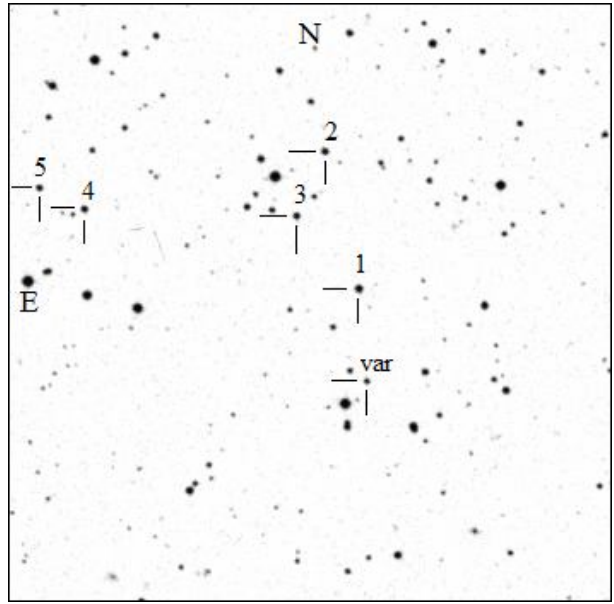

Fig. 2. Comparison stars for VSX J052807.9+725606.

Coordinates of center R.A. (2000)=$05^h28^m19^s$, Dec. (2000)=+72° 57″ 30′.

TABLE 3. Instrumental *bvr* magnitudes of comparison stars.

| No | USNO-B1.0 | R.A. | Dec. | b | v | r |
|---|---|---|---|---|---|---|
| 1 | 1629-0064823 | $05^h 8^m07.^s716$ | +72°57'39.25" | 15.590±0.025 | 14.890±0.024 | 14.433±0.020 |
| 2 | 1629-0064833 | $05^h28^m12.^s181$ | +72°59'59.14" | 16.070±0.023 | 15.400±0.022 | 14.939±0.023 |
| 3 | 1629-0064852 | $05^h28^m19.^s968$ | +72°58'57.88" | 16.536±0.031 | 15.688±0.022 | 15.085±0.023 |
| 4 | 1629-0064957 | $05^h29^m07.^s707$ | +72°59'24.70" | 15.993±0.025 | 15.378±0.022 | 14.977±0.020 |
| 5 | 1629-0064975 | $05^h29^m17.^s425$ | +72°59'50.29" | 16.209±0.027 | 15.609±0.027 | 15.231±0.019 |

TABLE 4. *BVRc* magnitudes of comparison stars reduced to the standard system.

| No | USNO-B1.0 | B | V | Rc | B-V | V-Rc |
|---|---|---|---|---|---|---|
| 1 | 1629-0064823 | 15.587±0.036 | 14.890±0.025 | 14.433±0.020 | 0.697±0.049 | 0.457±0.032 |
| 2 | 1629-0064833 | 16.055±0.034 | 15.401±0.023 | 14.940±0.022 | 0.654±0.046 | 0.461±0.032 |
| 3 | 1629-0064852 | 16.592±0.044 | 15.684±0.023 | 15.087±0.023 | 0.908±0.054 | 0.597±0.032 |
| 4 | 1629-0064957 | 15.955±0.037 | 15.380±0.022 | 14.977±0.020 | 0.575±0.048 | 0.404±0.030 |
| 5 | 1629-0064975 | 16.165±0.039 | 15.612±0.028 | 15.230±0.019 | 0.554±0.054 | 0.381±0.034 |

**Photometric results**

In [10] Virnina and Andronov indicated the photometric ephemerid of VSX J052807.9+725606:

$$Min.I.\ HJD = 2455261.8484 + 0.41179\ E$$
$$\pm0.0016\quad \pm0.00005$$

Our observations in photometric system *b, v, r* were approximated by the trigonometric polynomial with determining of the statistically optimal degree and rectification of the initial value of the period using the differential correction method routine. We used the FDCN software (Andronov [19, 20]).

For all three filters of the photometrical system we obtained the same value of the statistically optimal degree s=6, which coincides with the one obtained in [10] for the observations of the previous season. The "seasonal mean" values of period for *b, v* and *r* bands are $0.411725^d \pm 0.000014^d$, $0.411715^d \pm 0.000011^d$, $0.411736^d \pm 0.000011^d$. Within errors, these values coincide for all filters, thus the weighed value $P=0.411725^d \pm 0.000006^d$ has been calculated.

The minima timings are listed in the Table 5. The weighted initial epoch for the primary minimum is $T_0$=HJD 2455550.0374±0.0005. Besides, the individual minima timings for the primary and secondary minima were determined using the "asymptotic parabola" method [20, 21]. Whereas their accuracy is worse then the mean value, they could be used in future for constructing the *O-C* diagrams.

Using the mean value of the period from the previous observational season [10] and the current one, we obtained a corrected value $P=0.4116986^d \pm 0.0000024^d$. The difference in the periods for these two seasons $\Delta P$=-0.000065$^d$±0.0.000050$^d$ isn't statistically significant, thus for the studying of expected period changes (similar to V361 Lyr), it is necessary to continue monitoring of VSX J052807.9+725606.

Thus we used an ephemeris:

$$Min.I.\ HJD = 2455550.0374 + 0.4116986\ E,$$
$$\pm 0.0005\ \ \pm 0.0000024$$

The phase light curves in the instrumental photometric system *b, v, r* are shown on the Figure 3.

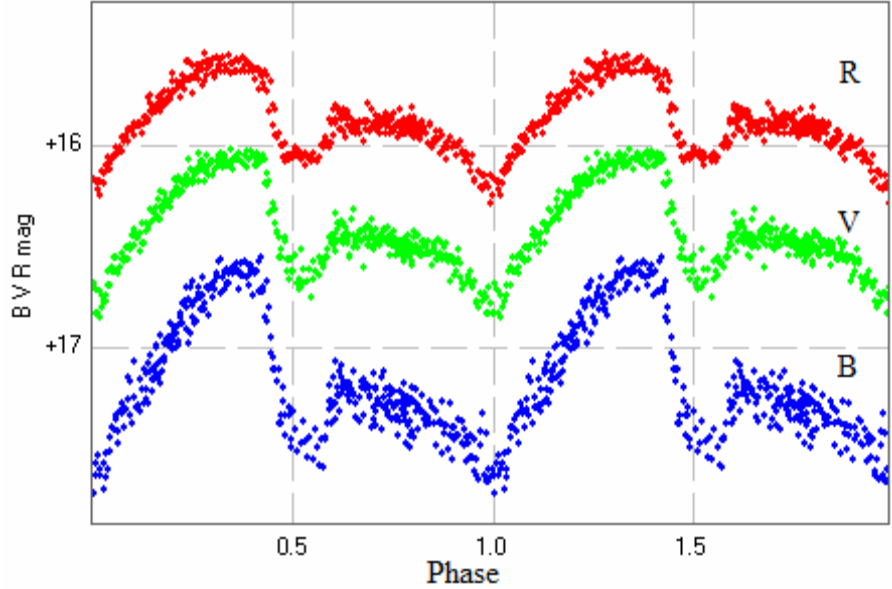

Fig 3. The phase light curves of VSX J052807.9+725606 in the instrumental systems *b, v* and *r*

Using the adopted value of the period, each phase curve had been smoothed by the statistically optimal trigonometric fit of degree *s,* using the least squares method routine, and the program MCV [22]. We also determined the stellar magnitudes in both maxima and minima in the instrumental bands b, v, r, and their phases. These magnitudes, phases and degrees of polynomials are listed in Table 4. The amplitudes *A* and the differences of magnitudes at maxima Δ*max* are given in the same table. The trigonometric polynomial approximations of brightness and dependence of accuracy of the smoothing curve on phase were transformed into standard *BVRc* systems, using the relations mentioned above.

Table 5. Timings of mean and individual (ind) minima of VSX J052807.9+725606 (HJD 24…..).

| *Rem* | *b* | *V* | *r* |
|---|---|---|---|
| Min I (mean) | 55550.0376±0.0014 | 55550.0369±0.0008 | 55550.0379±0.0008 |
| Min II (ind) | 55538.3212±0.0011 | 55538.3166±0.0016 | 55538.3157±0.0192 |
| Min I (ind) | 55539.3347±0.0033 | 55539.3358±0.0020 | 55539.3339±0.0027 |
| Min I (ind) | 55567.3421±0.0041 | 55567.3264±0.0023 | 55567.3321±0.0020 |

TABLE 6. Instrumental *b, v, r* magnitudes and magnitudes in the standard filters *B, V, Rc* in minima and maxima, degree of the trigonometric polynomial fits, full amplitude and the difference between the maxima.

| Parameters | *b* | *v* | *r* | *B* | *V* | *Rc* |
|---|---|---|---|---|---|---|
| $min_I$ $\phi=0.00$ | 17.595 ±0.015 | 16.785 ±0.008 | 16.186 ±0.008 | 17.656 ±0.021 | 16.781 ±0.015 | 16.187 ±0.007 |
| $min_{II}$ $\phi=0.53$ | 17.532 ±0.018 | 16.656 ±0.010 | 16.088 ±0.008 | 17.619 ±0.025 | 16.651 ±0.018 | 16.089 ±0.008 |
| $Max_I$ $\phi=0.38$ | 16.583 ±0.014 | 16.026 ±0.008 | 15.584 ±0.008 | 16.543 ±0.021 | 16.030 ±0.015 | 15.582 ±0.007 |
| $Max_{II}$ $\phi=0.64$ | 17.194 ±0.013 | 16.432 ±0.008 | 15.868 ±0.007 | 17.237 ±0.019 | 16.429 ±0.013 | 15.867 ±0.006 |
| s | 6 | 6 | 6 | 6 | 6 | 6 |
| A | 1.012 ±0.021 | 0.759 ±0.011 | 0.602 ±0.011 | 1.113 ±0.030 | 0.751 ±0.021 | 0.605 ±0.011 |
| Δmax | 0.611 ±0.018 | 0.406 ±0.011 | 0.284 ±0.011 | 0.694 ±0.028 | 0.399 ±0.020 | 0.286 ±0.010 |

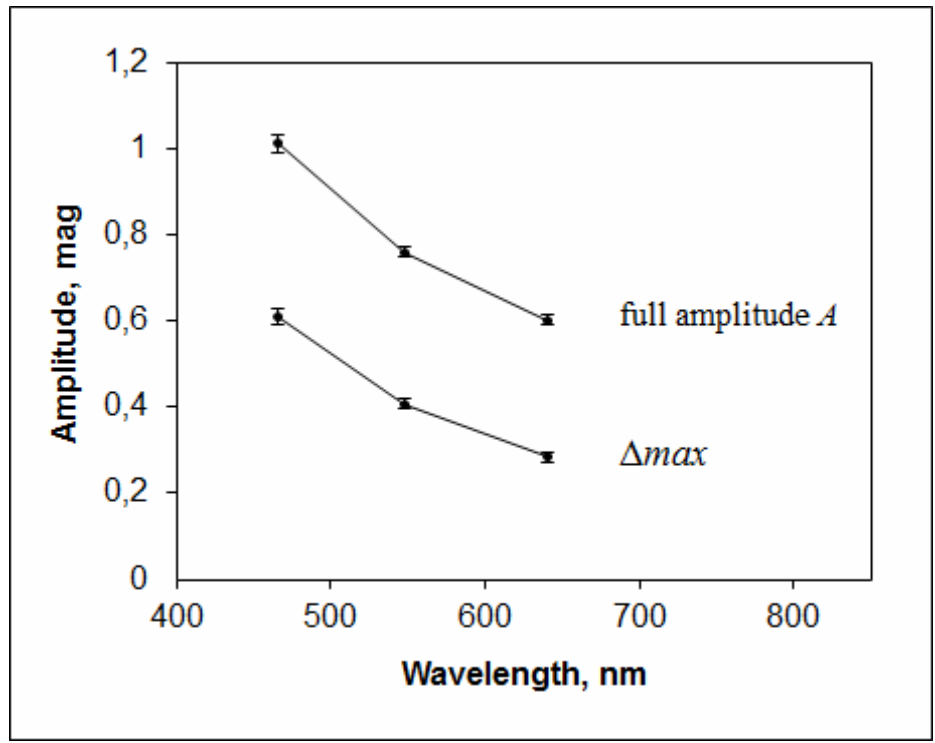

Fig 4. The wavelength dependence of full amplitudes and the difference between maxima, shown with error bars, for observations in the instrumental systems.

Like in the case of V361 Lyr, the amplitudes are different in different filters. The largest amplitude is observed in the *B*-band, the smallest is in *R*-band. We've got the same result for the difference in maxima: the longer wavelength the larger difference in magnitudes. These dependences are shown in the Table 5 and on the Figure 4.

**Color index analysis**

We obtained *B-V, V-Rc* and *B-Rc* color index phase curves, and smoothed the result by the trigonometric polynomial of degree *s*=6. The variations of the color indices are shown on the Figure 5.

The observed radiation from VSX J052807.9+725606 is the sum of emission from several sources with different temperatures: the primary component (eclipsed in the primary minimum), the secondary one and the "hot spot". There could also exist some groups of cool spots on one or both components, as there are suggested on the components of V361 Lyr.

As in the case of V361 Lyr, in the system VSX J052807.9+725606 the eclipses clearly aren't total; the color temperature, which could be estimated from the color indices, corresponds to the integral sum of emission from two or more sources.

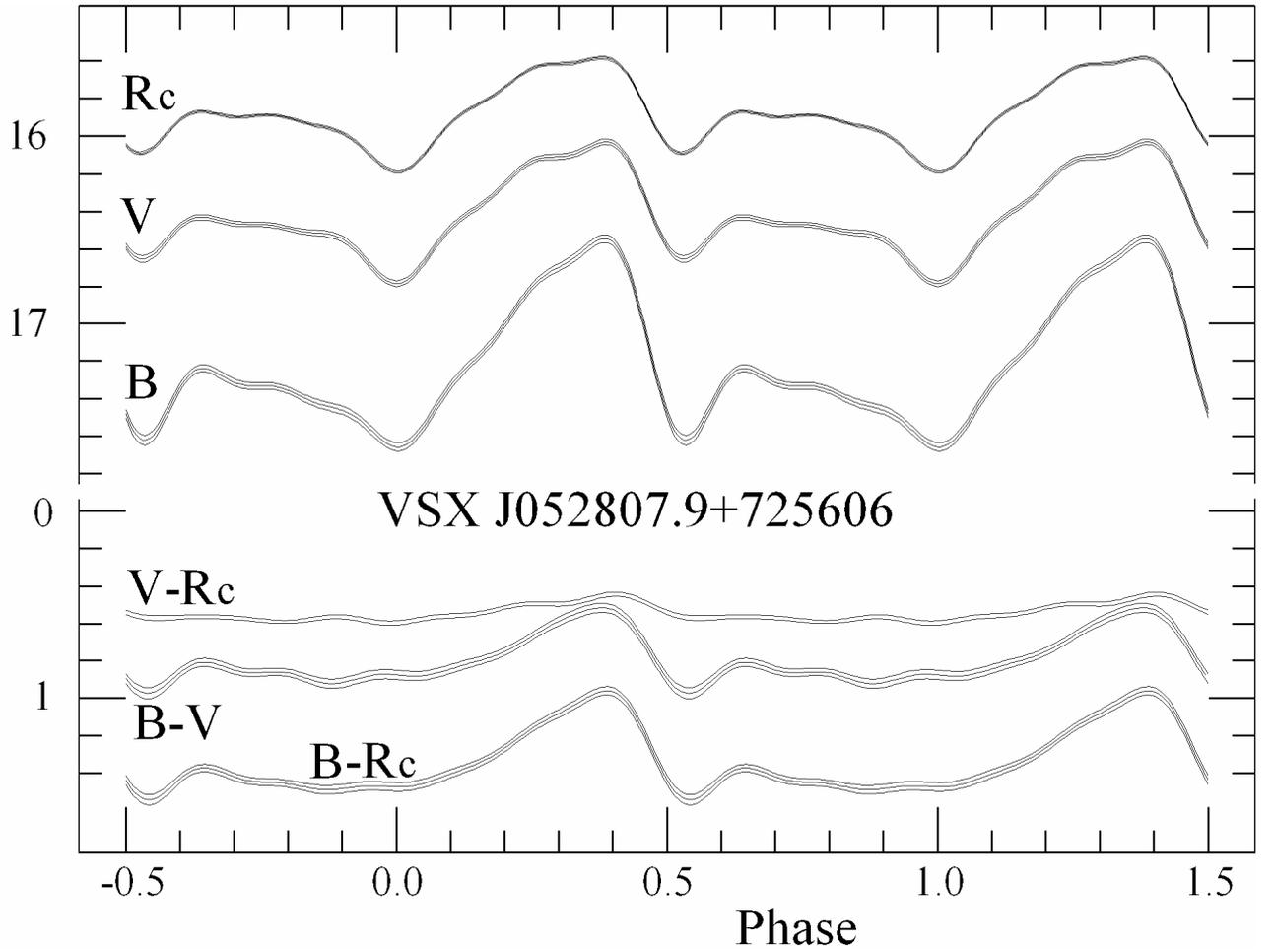

Fig 5. Smoothed phase curves of *B,V,Rc, B-V, V-Rc* and *B-Rc* color indices with $1\sigma$ error corridor.

To convert the color indices into temperatures T, we have used tables from [23] and [24] and used linear interpolation for the tabular dependence of $Q=10000/T$ on $(B-V)$.

From the smoothing curves, the *B-V* color indices in maxima are $(B-V)_{maxI}=0.513(\pm 0.026)$ *mag* and $(B-V)_{maxII}=0.808(\pm 0.023)$ *mag*, correspond to the color temperatures $6249(\pm 87)$ *K* and $5150(\pm 73)$ *K*, respectively; in primary and secondary minima $(B-V)_{minI}=0.875(\pm 0.026)$ *mag* and $(B-V)_{minII}=0.968(\pm 0.031)$ *mag*, which means that the respective color temperatures are $4943(\pm 60)$ *K* and $4666(\pm 88)$ *K*.

These color indices correspond to the source of emission with uniformly distributed temperature over the surface. However, as we interpret the primary (highest) maximum as a result of presence of a "hot spot" caused by "direct impact", we estimate its characteristics. For this purpose, we propose a model of "additional source"

$$I_{spot} = I_{Max\ I} - I_{Max\ II},$$

where $I=10^{-0.4m}$ is an intensity in arbitrary units, which corresponds to a stellar magnitude m, for the spot ($I_{spot}$), main ($I_{Max\ I}$) and secondary ($I_{Max\ II}$) maxima. Error estimates were determined using the formula $\sigma_I/I=0.4 \cdot \ln10 \cdot \sigma_m = 0.921\sigma_m$. The characteristics of the hot spot, determined for the standard $BVRc$ photometric system, are the following: $B=17.357^m \pm 0.049^m$, $V=17.310^m \pm 0.057^m$, $Rc=17.174^m \pm 0.036^m$, $B-V=0.047^m \pm 0.075^m$, $V-Rc=0.137^m \pm 0.067^m$. The color index $B-V$ corresponds to a temperature of the "hot spot" of $T=9341 \pm 766$ K, nearly twice larger than at the minima, when the major source of emission is one of the stars.

For these values of temperatures, the ratio of mean surface brightness of a "hot spot" and binary system (seen at the secondary maximum) is 17.8 and 10.3 for $B$ and $Vj$, respectively. Thus the ratio of the visible surface of the "hot spot" to that of the binary system is 0.050 (in $B$) and 0.043 (in $V$). This ~14% difference in estimates is not statistically significant for the error estimates listed above.

The temperatures and radii of both stellar components are unknown. The depth and shape of minima argue that there is no eclipse in the system, which could help to determine these parameters. Thus we make a very preliminary assumption that the sizes of both stars are comparable. The approximate estimate of the radius of the "hot spot" is ~30% of the radius of the star. This is roughly in an agreement with estimates for V361 Lyr.

Such estimates may be improved in accuracy after acquiring additional observations to decrease statistical errors, and as a result of modeling of the complete light curve using the programs taking into account spots on stars. However, results presented here may be used as initial values for further improvement of the model.

**Discussion**

We analyzed multicolor $BVR$ observations of an unusual newly discovered system VSX J052807.9+725606, obtained on 1.25 m telescope AZT-11 of Crimean Astrophysical Observatory. Using these observations, we calculated new ephemerides, which could be later used for the $O-C$ analysis. We confirmed a very strong asymmetry of the phase curve and found that, like in the case of V361 Lyr, the wavelength dependence of amplitude is observed. The maximum brightness corresponds to the phase 0.38, which is very close to that characteristic of V361 Lyr, and the secondary minimum is shifted from the phase 0.5. We also obtained the color temperatures in maxima and minima and found that the temperatures in minima are nearly equal.

Comparing VSX J052807.9+725606 and V361 Lyr, one can notice that these stars appeared to be "twins". However, several differences, which could be determined from photometrical studying, are present. The amplitude of V361 Lyr is higher than that one of VSX J052807.9+725606, the depths of minima are very close in the case of VSX J052807.9+725606, unlike at the phase curve of V361 Lyr. The temperatures of the components in V361 Lyr system are quiet different [8]. The color indices and the depths of minima show that the temperatures of the components of VSX J052807.9+725606 are relatively close.

In the temperature estimates, we have used internal accuracy of the smoothing curves. Another source of errors may be accuracy of calibration of comparison stars and an unknown interstellar extinction. Future possible corrections of the brightness and color index of comparison stars may slightly shift all the estimates quantitatively, but not qualitatively.

On the other hand, closeness of the phase curve's shapes and other photometrical characteristics of these two binary systems suggest the same physical processes, caused the asymmetry of the curve. Also both stars are short-periodical systems, their periods are close to each other.

From the *O-C* analysis of V361 Lyr ([7, 8, 9]), very rapid period decrease is clear. This star had been recognized as the binary system on its final stage of transition from detached configuration to contact one. The same effect we expect to detect for VSX J052807.9+725606. However, as this star is newly discovered and rather faint for any existing surveys, we have no archive data to enlarge the duration of observational series to check this guess. These facts, and the uniqueness of the group consisting of systems V361 Lyr and VSX J052807.9+725606 entail that this stage is a very short evolution phase of some binary systems, that's why the systems like these two are very rare and only two of them are found so far.

To prove the physical nature of VSX J052807.9+725606, further multicolor observations and spectral studying are desirable.

**Acknowledgments**. The authors are thankful to K.N.Grankin, V.P.Grinin, Marek Wolf and Zdeněk Mikulášek for helpful discussions.

**References**

1. G. A. Richter, I. L. Andronov, *Mitt. Veränderliche Sterne,* **11**, 27 (1986).
2. I. L. Andronov, G. A. Richter, Astronomische Nachrichten, **308**, 235 (1987).
3. J. Kaluzny, *Astron. J.*, **99**, 1207 (1990).
4. J. Kaluzny, Acta Astronomica, **41**, 17 (1991).
5. Di-Sheng Zhai, Ming-Jun Fang, *Chin. Astron. Astrophys.*, **19**, 334 (1995).
6. J. D. Gray, R. G. Samec, B. J. Carrigan, *Information Bulletin on Variable Stars,* **4177**, 1 (1995).
7. T. A. Lister, *COOL STARS, STELLAR SYSTEMS AND THE SUN: Proceedings of the 15th Cambridge Workshop on Cool Stars, Stellar Systems and the Sun. AIP Conference Proceedings*, **1094**, 688 (2009).
8. R.W. Hilditch, C.A. Collier, G. Hill, S.A. Bell,, T.J. Harries, *Mon. Notic. Roy. Astron. Soc.*, **291**, 749 (1997).
9. N. A. Virnina, *Open European Journal on Variable Stars*, **139**, 20 (2011)
10. N. A. Virnina, I. L. Andronov, *Open European Journal on Variable Stars*, **127**, 1 (2010).
11. O'Connell D. J. K., 1951, Pub. Riverview College Obs., 2, 85
12. N. A. Virnina, Odessa Astron. Publ., 23, 143 (2011)
13. J. Mergentaler, 1950, Wroclaw Contr., no. 4. p.1
14. L. Binnendijk, Astron.J., **65**, 358 (1960)
15. Qing-Yao Liu, Yu-Lan Yang, Chinese J. AsAp, 3, 142 (2003).
16. AAVSO variable stars plotter, http://www.aavso.org/vsp
17. H.Hardy, Reduction of Photoelectric Observations. –in: "Astronomical techniques", Stars and Stellar Systems, Chicago: University of Chicago Press, 1964, edited by W.A.Hiltner

18. A.U. Landolt, Astron. J. **88**, 439 (1983).
19. I. L. Andronov, *OAP*, **7**, 49 (1994).
20. I. L. Andronov, *ASPC*, **292**, 391 (2003).
21. V. I. Marsakova, I. L. Andronov, *OAP*, **9**, 127 (1996).
22. I. L. Andronov, A.V. Baklanov, Astronomy School Reports, **5**, 264 (2004), http://uavso.pochta.ru/mcv
23. C. W. Allen, *Astrophysical Quantities* [Russian translation], Mir, Moscow (1977).
24. V. Straizhys, *Multicolor stellar photometry. Photometric systems and methods* (in Russian), Mokslas, Vilnius (1977).